\documentclass{article}
\begin{document}
\title{Relativistic Lighthouses: The Role of the Binary Pulsar in
proving the existence of Gravitational Waves}
\author{Daniel Kennefick}
\maketitle

\section{Introduction}

In 1993 Joseph Taylor and Russel Hulse received the Nobel prize for their
discovery of the first binary pulsar, PSR 1913+16. Their citation acknowledged
the important of their work for the field of
gravitation and the accompanying press release stressed the special
significance of their measurements of the orbital motion of
the system in providing the first experimental evidence for the
existence of gravitational waves (Royal Swedish Academy of Sciences, 1993). 
Nobel prizes for work in astronomy and
astrophysics were once very rare, and prizes awarded for work in
gravitational physics have been even rarer. Einstein himself was denied 
any citation for his discovery of General Relativity. This Nobel prize
is therefore a striking
demonstration of the importance of gravitational waves as a
topic of physics. It is all the more interesting then, that gravitational
wave research had been, until Taylor and Hulse's discovery, a marginal and
controversial field. The very existence of gravitational waves, even
as a theoretical prediction of Einstein's theory, had frequently been
doubted in the half century before Taylor and Hulse's work. The story of 
how this classic experiment by two astrophysicists
settled a long-standing theoretical controversy seems a natural
for study from a historical perspective. 

My own interest in the binary pulsar experiment derives primarily
from its relation to the long running quadrupole formula controversy
which centered on the validity, within general relativity, 
of the quadrupole formula, first derived by Einstein in 1918, when
applied to binary star systems such as the binary pulsar (Kennefick 2007).
This aspect of the binary pulsar's history, while interesting in itself,
is highly relevant to the question of
proving the existence of gravitational waves. What Taylor and Hulse 
achieved was to show that the rate of decay of the binary pulsar's orbit
was in agreement with the prediction of the quadrupole formula, suggesting
emission of gravitational radiation by the system as the cause of the decay.
Obviously this line of logic depended on the assumption that the quadrupole
formula was correctly derived from the established theory. 
\footnote{An interesting side note to the binary pulsar discovery of gravitational
waves is the common use of phrases like ``indirect detection'' of gravity
waves, to distinguish Taylor and company's work from the long awaited
``direct detection'' of gravitational waves by Earth-based detectors. A
number of people have pointed out the fallacy in this kind of thinking,
observing that, strictly speaking, the binary pulsar evidence is no
more indirect than any other detection. Two people I am thinking
of particularly are Thibault Damour and Allan Franklin, both of whom have
made the point to me personally. While it is true that the 
astronomers use electromagnetic signals from the source system, and
must then infer the presence of gravitational waves from the observed
behavior, the same is true of Earth-based detectors, which also use
electromagnetically controlled detection of local masses and deduce
the presence of gravitational waves from the motions of those masses.
In some sense the only difference is that the binary pulsar
astronomers only observe the source of the gravitational waves, and
thus cannot comment on the propagation through space of these waves.
Another distinction, of some relevance to our discussion, is that
the theory required to analyze the binary pulsar system is not the
linearized gravity which suffices for the Earth-based detector, and
therefore, it could be argued, it is a more complex and more
controversial process of deduction. This claim would be, however,
highly debatable. To date, the experimental
detection of gravitational waves has been even more controversial
than the theoretical study of their sources, and there have been
several challenges over the years to the established theory of
such detectors.}
This comes even more into 
focus when one realizes that the theoretical controversy over gravitational
waves, in its early phases, focused on the question of whether
gravitational waves could exist at all, or could be emitted by binary stars.

This topic bears somewhat
on the longstanding debate between Harry Collins and Allan Franklin
over Collins' concept of the Experimenters' Regress, as applied to
the Weber controversy, coincidentally enough a controversy over the 
detection of gravitational waves by Earth-based detectors. 
It is well known that the exchange between Collins 
and Franklin over the Experimenters' Regress and Collins' interpretation
of the controversy over gravitational wave detection
was seen as a significant engagement in the so-called Science Wars.
At that time philosophers, like Franklin, debated with sociologists, 
like Collins, over issues such as realism versus relativism, the demarcation
problem in science studies and so on. The debate between Collins and Franklin
centered on whether physicists had rational or objective grounds for
closing debates, or whether achieving closure in scientific controversies
depended on the social relations between the participants and their
community. Sociologists like Collins
argue that ``interpretive flexibility'' means that physicists always
have the option to keep a debate open, but that social cohesion depends
upon the ability of the core group to eventually achieve a consensus, even
in defiance of the wishes of the remaining group of outsiders
who regard the matter as unsettled (Collins 1994, Collins 2004). 
Franklin's take was that the rump group
of outsiders in the Weber controversy were behaving irrationally in refusing
to accept comprehensive experimental evidence contradicting their view
and had, in some sense, placed
themselves outside of the sphere of rational scientific discourse (Franklin
, 1994).

Inspired
by Collins' work, I made use of the analogous concept of the Theoreticians'
Regress to explain the intractability of the controversy on the
theoretical side of the gravitational wave field. Somewhat to my relief,
my own study did not appear to be so controversial, in the context of
the science wars, no doubt largely because of my insignificant status in
the field. Additionally,
neither side claimed that theorists were directly confronted with the
objective reality of the laboratory. But what about the fact
that the close of the debate in my story was apparently connected,
certainly timed so as to suggest a connection, with the arrival of
experimental evidence? Was it not particularly satisfactory for
``realists'' that unambiguous, and largely unchallenged, experimental
evidence should help to close out debate amongst theorists? While I was
happy that my study was not likely to play a role in the science
wars, I was shy of this one issue that was apparently
relevant to the questions at issue in that struggle.

Both Collins and Franklin were pioneers in the careful micro-study
of experimental method and practice. Through personal contact with both
men, and others, I was inspired to adopt this kind of approach. From 
my perspective
Collins and Franklin appeared to be saying very
similar things. I confess to being somewhat uncomfortable in addressing
the precise role played by the binary pulsar in settling the quadrupole
formula controversy, lest I be seen to be firing a shot in a war which,
however interesting the individual debates, I find slightly incomprehensible.
Surely what mattered was that both Collins and Franklin believed in close
detailed studies of what scientists actually did, not in whether they
agreed on points of principal? Here, I suppose, my own outlook as a historian
differed from either Collins or Franklin, who as a sociologist and
a philosopher, respectively, viewed the micro-study as a means to an
end. Their goal lies not only in the interest of the study itself,
but also in what it teaches us about the way science is done.
At any rate, it seemed as if I was a conscientious objector in the science wars.
Now that a ceasefire amongst those who value the scientific endeavor
seems likely to endure (Collins 2009)
this paper is by way of being a belated commentary on this issue, from the
perspective of a draft dodger now returned to the scene of the 
fray in peacetime.

\section{Controversy}

The background to the story can be sketched relatively briefly (for
a fuller account, see Kennefick 2007). The theory of gravitational waves
dates to 1916 with Einstein's first paper on the subject, only half a year
after his publication of the final form of his general relativity theory. 
In 1918 Einstein
published a paper correcting a certain error from the paper of 1916, and
presenting, for the first time, the quadrupole formula, expressing the
rate of emission of gravitational wave energy by a system of accelerating
masses.

When Einstein derived the quadrupole formula it was on the basis of the
linearized approximation of general relativity. This permitted him to
make the calculation relatively straightforward, because in the coordinate
system adopted by him the linearized equations of gravity take on a form
which is directly analogous to the Maxwell equations for electromagnetism,
a theory in which the role of radiation was, and is, reasonably well 
understood.
But, since general relativity is a non-linear theory, this linearized
approximation can hold only for very weak fields, which specifically 
excludes systems, such as a binary star system, which are held together
by their own gravitational interaction. Since it is only this type of
system which (as far as we know today) might be capable of producing
detectable gravitational waves, this approximation leaves something to be
desired as far as sources go. It is thought to be ideal for the study of 
gravitational
wave detectors however. The question then is, does the quadrupole formula
give a reasonable approximation of the source strength of possible 
astrophysical sources of gravitational waves, especially binary stars?

Famously, Einstein himself came to entertain doubts about the existence
of gravitational waves (indeed, there is evidence that his paper of 1916
was preceded by a brief period of skepticism on the subject,
see Kennefick 2007, pp. 44-49), when he and his then assistant Nathan Rosen came to
look for an exact solution of the Einstein equations representing plane 
gravitational
waves (Einstein and Rosen, 1937). 
They discovered that it was not possible to construct a metric
in a given coordinate system which did not include a singularity somewhere
in the spacetime representing the plane gravitational waves. Subsequently
it was shown that this singularity is merely a coordinate singularity,
rather than a physical singularity, but at the time Einstein and Rosen
interpreted it as physical, arguing that such spacetimes could not exist.
However, before the paper was published, Einstein realized that his
argument was mistaken. Nevertheless, in the published version, he still
included a discussion of the possibility that binary stars would not
emit gravitational waves,
in spite of the fact that the quadrupole formula
suggests that they would. Einstein's assistant who succeeded Rosen,
Leopold Infeld afterwards always insisted that this was
Einstein's final word on the subject which, in a strictly published sense,
it was.

When interest in general relativity began to pick up again in the mid-fifties, Rosen and
Infeld advanced a number of arguments whose common point was that 
binary star systems would not undergo orbital decay as a result of
emitting gravitational waves. Hermann Bondi also entertained serious
doubts on this score, arguing that the analogy with electromagnetism
which lay behind the original notion of gravitational waves, actually
pointed this way. His view was that in electrodynamics it was believed
that accelerating charges emitted radiation and that the same was
expected to hold true in the case of gravity. But since the theory was
a theory of general relativity, how did one define what was accelerating?
In Bondi's view, an inertial particle in general relativity was one which followed a 
geodesic. An accelerating particle was one which did not. Since binary
stars in orbit around each other followed the geodesics of the local
spacetime, they were not accelerating, in this sense. As particles in
a form of inertial motion, their motion would not be of the type which
should decay in response to radiation reaction.

These kinds of arguments 
came up for discussion at a seminal 1957 meeting at Chapel Hill, North
Carolina, which was the inspiration for the General Relativity and 
Gravitation series of meeting which
have continued to the present day as the leading conferences in the field
of general relativity. The meeting is important for the history of
gravitational waves because it was there, in response to arguments raised
by Rosen, that Richard Feynman and Bondi himself,
responding to the work of Felix Pirani, put forward the ``sticky bead''
argument that gravitational waves must carry energy. As a result of this,
the debate shifted to the question of whether binary stars could emit
gravitational waves. This question was still being debated at the 
third General Relativity and Gravitation meeting
held in Warsaw in 1962. Feynman attended this meeting
and, one may speculate, was perturbed to find that the questions he
had thought were settled in 1957 were still being aired. While the questions
had changed somewhat, nevertheless Feynman had, in 1957 made an impassioned
case for the field to abandon a ``too rigorous'' approach as being infertile
in theoretical physics (De Witt, 1957). In a celebrated letter home from the 
conference to his wife, Feynman painted a Felliniesque portrait of a physicist
trapped inside a field full of ``dopes'' (126 of them at 
the conference,
according to his letter) rehearsing the same
arguments over and over again like ``a lot of worms trying to get out
of a bottle by crawling all over each other.'' (Feynman and Leighton, 1989)

Ironically enough, it was the work of Bondi himself, as much as of
any other relativist, which did the most to convince most relativists
that binary stars did indeed decay in their orbits as a result of
gravitational wave emission. But the debate seemed of little practical
relevance, since the one thing that everyone involved agreed upon
was that the rate at which this decay took place was too small for
it to be observable in any known orbital system. Very likely it was
for this reason that the debate became very quiet in the decade between
1965 and 1975. The discovery of the binary pulsar in late 1974 
undoubtedly did much to reinvigorate this debate, which by then had
shifted to a new question, whether the quadrupole formula was the correct
formula for strong gravity binaries of this kind. Over the course of
the following decade the debate was fairly  vigorous, until it petered out
in the mid-1980s, when the remaining skeptics grew quiet (again, for
a discussion of all of this history, with references, consult Kennefick
2007).

What is interesting about the role of the binary pulsar in this story
is that there are good grounds for believing that its primary role was
to stimulate the controversy into new life. It is usually thought
of as the agency by which the controversy was settled (and this is
certainly a role which is of interest to this paper), but another 
possible reading is that it actually made the controversy more 
prominent and more contentious and that this served, with time, to 
bring it to a conclusion by focusing the attention of theorists upon
it. One might speculate that we are dealing with a controversy 
downsizing principle, in 
analogy with the problem of cosmic downsizing in 
extragalactic astronomy, which revolves around the observation that 
over time quasars come to
have smaller and smaller black holes. Since black holes should only
every grow in size, it is claimed that this observational effect
arises because the big ones have already
used up all their fuel and ``turned off.'' The situation is thought to
be similar to that which obtains for stars, where the larger stars,
which paradoxically contain more fuel, burn the fuel at a far faster
rate and live a much shorter life than do less massive stars. 

In the case of scientific controversies we may similarly expect, at
any given moment to find many more small and almost moribund controversies
than strident ones, because the former will be more long-lived. 
The fuel which is only slowly consumed in a small controversy
is not the number of
issues to be debated. I agree with those who think such points
are all but inexhaustible. The fuel is the number of
potential participants in the controversy. Where the number of participants
is low, each of them may feel comfortable conceding a long period
of debate to what is a manageable number of colleagues. As the
number involved in the controversy rises, the ability to mediate
the controversy by direct personal relations between all participants
is strained. The consequences of remaining on the fence become less
predictable as they become potentially more serious, since more
people involved means potentially more influential people having a
vested interest in the outcome.
The participants come under pressure to take a definitive position
and tend to do so more quickly. To continue with the analogy,
the fuel is more quickly processed through the various stages,
from open minded participant, to committed protagonist, to close-minded
ideologue, at the end of which no further debate is possible. In essence,
the controversy which burns most brightly extinguishes itself most
quickly. To be sure, I am here merely taking a long-established piece
of folk wisdom and dressing it up in academic clothes. The phrase
``slow-burning controversy,'' already nicely encapsulates the image
I am trying to convey.

So let us examine briefly the course of the quadrupole formula controversy
in the 1970s. We have already summarized the debate over whether binary
stars could emit gravitational waves, a debate which flourished
in the late fifties and early sixties. There then followed a period in
which it was regarded as settled, by a large majority, that binary stars
did undergo radiation damping as a result of gravitational wave emission.
The detail of how this occurred was perhaps not regarded as a terribly
pressing problem, given that no one was familiar with any known astronomical
systems which, according to the quadrupole formula itself, would undergo
a measurable decay in their orbits. The state of affairs bore a close
approximation to the situation in controversies which have passed the
point of crystallization, which is to say that even though there remained
some who doubted the consensus opinion that the quadrupole formula was
approximately correct, their views did not receive much public airing.
In fact, however, it was still possible for their views to be aired,
the problem was simply not important enough for huge notice to be taken
of anyone's views on the matter.

A good example of the status of the debate on the eve of the discovery
of the binary pulsar is the June, 1973 Paris meeting on gravitational waves
at which Havas gave a talk outlining his view that the question whether
binary stars did emit gravitational waves at all was still unsettled,
and advancing his critique of the main calculations which agreed with the
quadrupole formula result (Havas, 1973). In the conference proceedings, 
two of the
remarks in response to Havas' talk can be regarded as sharing his
skepticism, two as disagreeing with it, and two as neutral (at least
phrased in a neutral way). This certainly suggests not only that Havas
had leave to raise such issues with his peers, but also that he
had an audience part of which, at least, was sympathetic. At the same
time, the problem was not at the forefront of theoretical concerns
at that moment. It was not considered irrelevant or uninteresting,
after all the very fact of the conference being held at all suggests
otherwise,
but the fact that no astrophysical applications had been discovered
certainly restricted its urgency.

Within little over a year the situation was transformed completely.

\section{Discovery}

Pulsars were discovered in 1967 by Jocelyn Bell and Tony Hewish 
using the Interplanetary Scintillation Array at 
the Mullard Radio Astronomy Observatory near Cambridge, England. 
It quickly became apparent that
pulsars were a real-life instance of a long standing theoretical entity,
the neutron star, which had been first proposed 
by Walter Baade and Fritz Zwicky decades previously, in 1933 (see
Haensel, Potekhin and Yakovlev, 2007, pp. 2-4 for a brief
history). The problem of gravitationally collapsed
objects become of greater theoretical interest
following the discovery of quasars by radio astronomers in the fifties
and was further stimulated by the pulsar discovery.
By the early seventies only a few dozen pulsars were known, and Joe
Taylor of the University of Massachusetts, together with his graduate
student Russell Hulse, proposed to do a computerized search for them
with the large Arecibo dish in Puerto Rico to provide a much larger
ensemble of discovered objects. It was a specific aim of Taylor's
proposal that such a large number of pulsars might feature one
which was part of a binary system (Hulse 1997). This would permit
the measurement of the mass of the pulsar, a topic of immense
astrophysical interest, since the very idea of neutron stars had
arisen following the work of Subramanian Chandrasekhar on the limiting
mass of white dwarf stars. That a close binary neutron star system
had been suggested as a possible source of detectable gravitational
waves as early as 1963 by Freeman Dyson was almost certainly not
on Taylor's mind as he began his pulsar search. This was all the more
true since Dyson's suggestion had been made in the context of a
suggestion that arbitrarily advanced alien civilizations might construct
such systems for the purpose of interstellar navigation.

In early July 1974 Hulse, down at Arecibo, recorded a pulsar, just
barely strong enough to be detected by the system, unusually sensitive
for its day as it was, whose position on the sky automatically
baptized it with the name PSR 1913+16. After confirmation that this
was indeed a pulsar, including measuring its period, Hulse recorded
the word ``fantastic'' on his observing record, referring to the fact
that the pulsar had the second shortest period known at that time. At
this point he had no notion that it was in a binary system, only
the rotational period of the neutron star itself had been measured,
not its orbital period. The only foretaste of what was to come was
that subsequent attempts to confirm that rapid pulse in these first
observations did not agree, to Hulse's frustration. He even went so
far as to cross out and erase these subsequent attempts from his log
(Hulse 1997).

In late August Hulse returned to this object, in a routine way, to
try to confirm its period. As before he found that its period kept
changing with each measurement. Indeed, by a curious coincidence,
he found that he almost repeated the same set of measurements each
time the pulsar came overhead at Arecibo (the dish at Arecibo is
so large it is built into a small valley, and thus cannot observe
very far from the zenith of the sky). This would turn out to be due
to the fact that the pulsar binary has an orbital period of just
under 8 hours, and thus completes a little over 3 orbits with every 
rotation of the Earth. It did not take Hulse long to convince himself
that he had discovered a pulsar in a binary system, and it was
immediately clear to him and to his advisor Taylor that they were
dealing with an extraordinary system. An eight hour orbital period
represented an orbiting system involving massive objects with
an unprecedently small physical separation from each other. 
Indeed word got around quickly about the new discovery, to the extent
that the first theoretical paper commenting on the binary pulsar appeared
in late 1974 (Damour and Ruffini, 1974), 
while the discovery paper itself, by Hulse and Taylor, appeared only in 1975.

There can be little doubt that interest in the radiation problem from
binary stars was reinvigorated by the binary pulsar discovery. Here was
a real world example of a system where radiation damping might actually
be measurable. Of course there were doubts expressed, on the theoretical
side (Damour and Ruffini 1974) that the effect really would be measurable,
but the experimenters were nevertheless not ruling it out. In an interview
Joe Taylor recalls his own view at the time (interview conducted by
the author by phone on 2nd May, 2008) ...

\begin{quotation}
The person who put us onto that was 
Bob Wagoner. It happened that once the news 
was out and it became public that this thing was there and that we were 
observing it, I responded to a number of invitations to go and give talks 
about it and ended up making a grand tour around North America 
where I made five or six stops and one of them was at Stanford and Bob 
Wagoner there actually gave me his paper predicting the orbital period 
decay to carry back with me since he knew I was going to be at Harvard a 
couple of days later and I handed it to Alex Dalgarno the editor of ApJ 
Letters. So it was Bob's paper [Wagoner 1975] that I first 
began to take seriously and to 
recognize that with the current state of the art then, in October 1974  
of doing pulsar timing, it was clear that, if his 
numbers were right, and I assumed they were, it would take us a number of years 
to see any effect, but not an unreasonable number and if we could improve 
the timing accuracy a little bit it might happen even sooner and that's 
more or less what happened.
\end{quotation}

While relativists were excited about a number of tests of general
relativity which could be made for this system whose components were
moving under the influence of unprecedently strong gravitational
forces, it seems that the measurement of the binary
pulsar orbital decay came significantly earlier than most people expected,
as Taylor agrees (interview, 2nd May, 2008):

\begin{quotation}
I think that's right and that's largely because at that time it 
wasn't yet recognized that doing really high precision timing of pulsar 
signals was a very important goal.
\end{quotation}

Nevertheless the possibility was in the air from late 1974 onwards,
and the fact that it would take a significant amount of time gave the
theorists ample time in which to apply new techniques and increased
effort to the problem of analyzing the orbital evolution of such a
system as it responded to its own gravitational wave emission.

To what extent was this activity on the theoretical side visible to
the experimenters? Given that their result, when available, was likely
to have a decisive effect on the controversy, it is remarkable that
they went totally unaware of it until they finally had a result to
announce. This announcement was made, in its earliest version, at the
ninth Texas Symposium on Relativistic Astrophysics in Munich in 1978.
The Texas series of meetings had a tradition of announcements of
important observational results. The first Texas meeting had been held in
response to the growing interest in quasars as new objects discovered
by radio astronomers in the late fifties.
Taylor's talk in Munich is one of the more celebrated of the
announcements made at this series of meetings (interview, 2nd May, 2008).

\begin{quotation}
Well, I'll tell you when I first even knew that there was any debate,
was at the Texas Symposium in Munich.\footnote{At this point on the
interview recording, the author can hear himself say `Really.'} And so somebody 
asked me a question, well let me back up just a little bit. I was scheduled 
to give a paper there on something like the second or third day of the 
conference, and J\"{u}rgen Ehlers, who was one of the conference organizers, 
recognized that somehow not getting to this until nearly the last day of the 
conference was not a good idea. So he asked me to get up and say just a few 
words about it in a session on the first day so that at least people would 
know what I looked like and we could talk in the halls, and so forth, 
afterwards. So I did that and I basically gave the result and said I'll give 
all the details at the scheduled time the day after tomorrow, or something 
like that. Somebody then in the audience asked a question, I don't remember 
who it was, `when you say that you have seen the period decay and it 
agrees with the prediction, what prediction are you using?' And I sort of was 
blind-sided by that. I just thought that everyone knew how to calculate this, 
except maybe me. And so I think I must have stood there wondering how to 
answer for a minute and Tommy Gold, who happened to be the session chairman, 
whispered in my ear, `Landau and Lifshitz,' so I said it's given in Landau and 
Lifshitz. So that more or less is what transpired. I mean, I remember having 
conversations later with people about it and I began to realize that, of 
course, that was just sort of an heuristic formula and the calculation wasn't 
even derived, I guess, in Landau and Lifshitz, it was given as an exercise for 
the student to do. 
\end{quotation}

It is humorous to note that Gold, the session chairman, had been, with
his collaborator Bondi, one
of the early skeptics concerning whether binary stars could emit
gravitational radiation. Although Gold would certainly have been very
familiar with Landau and Lifshitz' treatment, he might also have been
inclined to agree with Bondi's comment, that it was very ``glib.''

So once Taylor was apprised of the existence of the controversy, what
was his reaction (interview, 2nd May, 2008)?

\begin{quotation}
So ok, so I was aware then that there was a controversy about it. Whenever 
I quizzed theorists, that I knew pretty well, about it, 
they tended to be people 
like Kip Thorne, for example. Kip always said, `oh yes, you know, we're still 
worrying about the mathematical details, but we know its right.' And my 
impression was that, I think pretty much I gained the impression that you 
convey to a large extent in your book as well,\footnote{A reference to 
Kennefick 2007, illustrating one of the problems faced by an oral historian
who wishes to write books and continue doing oral histories!} that the more 
mathematically oriented physicists, and particularly those who had been doing 
relativity in mathematics departments, were still concerned about the lack of 
rigor and the full mathematical beauty, but the physicists like 
Thorne and Feynman and others just had little patience with that kind 
of concern and wanted to get on with it and see what you could do with it. 
And they more or less told me `don't worry about it.'
\end{quotation}

So communication between theorists and experimenters contained this
interesting feature, that a reasonably lively controversy 
amongst the theorists could be completely invisible to the experimenters. 
Obviously the controversy was
not one which consumed the total energy of theorists in the
field, but it still involved a good deal of back and forth and even
a dedicated workshop, during the period in question, and yet no mention
was made of its existence within Taylor's hearing. 
Partly, as Taylor
says, this was because of the kind of theorists he was talking to.
In the field of relativistic astrophysics, there were people close
to the astrophysics end of the spectrum, and people closer to the
relativity end, and Taylor, as an astrophysicist, was naturally more
likely to talk to those on the astrophysics end. Since those theorists
were less likely to be skeptical of the quadrupole formula, they
naturally chose not to bring up any caveats about the derivations
which they felt were unlikely ever to have a bearing on the
observations underway. Furthermore, and this bears on a point
I will try to bring out at the end of the paper, they may have felt
some slight embarrassment that there existed theorists in their field
who still doubted the canonical understanding of gravitational radiation
in general relativity.

\section{Trading Zones and Pidgins}

In his book {\it Image and Logic} Peter Galison (1997), another pioneer
of the careful micro-study of physicists in action, argues that different
groups of scientists, in particular experimental and theoretical physicists
often speak different technical languages and encounter difficulty in
communicating with each other. He argues that, in such situations, 
physicists find it useful to develop a pidgin, a term used to describe
a secondary language, formed usually from a mishmash of other languages,
used to facilitate trade between different peoples. Galison describes
the conceptual space between different groups of physicists as a trading
zone and discusses the use of pidgins, which in his usage may refer to
particular mathematical constructs designed to permit experimenters and
theoreticians (let's say) to discuss and compare the 
predictions of the latter with the results of the former. 

The binary pulsar is an interesting case to observe the possible
need for trading zones, since it was a discovery by
radio astronomers who had, otherwise, relatively little contact with
relativists interested in gravitational waves. At the same time 
their field had arisen alongside the broader culture of relativistic
astrophysics, which was formed by a first contact between radio astronomers
and relativists after the discovery of quasars.
To what extent do we observe the need for a trading zone between
experimenters and theorists in our particular story? Certainly there
seem to be areas of physics in which theorists and experimenters talk
to each other regularly and apparently freely, and it is certainly also
true that when physicists, even from very different subject areas, converse,
they speak a recognizable technical language which seems to be quite
unconscious of boundaries. Indeed, for the physicist, the international,
inter subject quality of physics speech is one of the defining experiences
of being a physicist (no doubt the same may be true for scholars in other
disciplines). Nevertheless there is some evidence, in the case of the
binary pulsar story, supporting the model put forward by Galison.
One promising
way to understand how scientists deal with trading zones, when and if
they occur, is through the notion of {\it interactional expertise},
a concept which describes the ability of someone to talk intelligibly
and usefully to an expert about their field, even if they are not (yet)
capable of working in that field, which would be full expertise (Collins,
Evans and Gorman, 2007). It may be that, even where physicists lack
direct expertise to work in a neighboring field, they at least possess
interactional expertise to talk with their fellow physicists in that
field.

Let us begin with the discovery of the binary pulsar in 1974. The two
astronomers involved, Joseph Taylor and Russell Hulse, both received
educations fairly typical of astronomers of their generation in that
they were educated primarily in physics (in fact
Hulse was still a graduate student when he discovered the binary pulsar).
In this context, particularly as the two men were working in radio
astronomy, astronomy is conceived
of as being more or less a sub-discipline of physics, albeit an
unusually ancient one which still maintained a certain level of
institutional independence. As such they took courses in general
relativity, a subject within physics which was typically considered
an optional higher level course, but one which might be especially
relevant to those planning to specialize in astronomy. As radio 
astronomers interested in pulsars, relativity theory was clearly
relevant to an understanding of the source of the signals they
planned to study, but not nearly as relevant and routine as the
physics of the electromagnetically based detectors and instruments
they operated. 

Accordingly Joe Taylor describes one of his first actions on 
discovering that he had a binary pulsar with a uniquely close orbit
involving unprecedently intense gravitational interaction between
the two components (interview, 2nd May, 2008). 

\begin{quotation}
We'd both taken the obligatory,
or almost obligatory, relativity course in University, as part of our 
physics training, but neither one of us was very deeply into relativity. 
My wife was much amused when one day, this was when I was at the 
University of Massachusetts, of course, I said I don't have to teach 
today, I'm going to drive into Boston and visit the Tech Coop. And I 
spent the day in the MIT bookstore and came back with a pile of books, 
Weinberg, and Misner, Thorne and Wheeler and all the other ones that you 
would imagine. She was much amused that I spent the next few months deeply 
engrossed in these books.
\end{quotation}

So certainly the astronomers felt a need to get up to speed with the
elements of relativistic orbital motion. To what extent was there a
language gap between them and the practitioners of this discipline?
Partly the gap was a social gap. Neither Taylor nor Hulse habituated
amongst relativists and therefore did not partake in their discourse.
So Taylor went unaware of the ongoing
quadrupole formula controversy, throughout the time when, as we would
be tempted to say today, he was determining the outcome of this controversy.

But leaving aside this question of discourse, when Taylor and his
collaborators did speak to relativists, could they make themselves
understood and be understood? Clearly they could, for the most part.
But some obstacles were encountered. By the time Taylor and company
were dealing with the orbital decay of the binary pulsar, Hulse had
finished his doctorate and moved on. A collaborator with whom Taylor
published many of the early papers announcing and discussing
the orbital decay was Joel Weisberg.
Weisberg does recall language difficulty playing some modest role in
talking to theorists, before they found a long term collaborator
in a talented young French relativist, Thibault Damour (interview
conducted by the author, by phone, on 24th February, 2000).

\begin{quotation}
It's interesting, we had a failed attempt to work with one person. And I 
think the problem was he couldn't talk well enough to experimentalists. 
He couldn't give us results that were easily interpretable by us, 
whereas Thibault could. It was quite interesting.
\end{quotation}

Weisberg describes the kind of theorist that would be helpful in
the process of theory testing using the binary pulsar data, saying
``it had to be people who could talk a language I could understand.''  
Regarding the one failed effort mentioned above, the problem had
a very practical aspect, 
``he [the theorist] couldn't give us specific things to test.''
At the same time he emphasizes that their eventual collaborator, Damour
was ``brilliant'' and ``made fundamental progress,'' so 
``it wasn't just a language thing.'' He adds (in a private communication)
that the ``theorist `speaking the right language' was not, by itself,
enough for a successful collaboration.''

Nevertheless, to examine the ``language thing,'' I suspect it is
fair to say that, in the absence of a relativity community, Taylor
and Weisberg would have been capable of performing calculations
to establish the predictions of certain theories. In fact, as
we shall see, they did contribute original work on the theory
side. The problem seems
to me to be legitimately a question of language and society, in 
the sense that Taylor and Weisberg's problem was not primarily that
they lacked the expertise to do the calculations. That much they
could have acquired, and did acquire, with time and effort. What
they lacked was fluency in the language spoken by theorists, and
social standing within the discourse of theory. The existence of theories
to test is inextricably linked with the existence of theorists
who developed them, who have a vested interest in the testing. 
Since the theorists are the experts, it is understandable that
the astronomers, like Taylor and Weisberg, would feel distinctly
hesitant about publicly putting forth calculations in an area that was not
their own realm of expertise. At the same time it was important
that the calculations which were done by theorists were not black
boxes whose inner workings were totally opaque to the experimenters.
It was important that the results of these calculations could be couched in
a form which dealt with observables pertinent to the actual
measurements being made.

The need for what Galison would describe as a pidgin seems to have
produced the parameterized post-Newtonian (PPN) framework as a tool
to mediate the theory testing process. This process required an alliance
of theorists and experimenters. Theorists made predictions based on
their calculations. Experimenters made measurements which were then
compared to the results of the calculations. 
This PPN framework had been widely used
during solar system tests of general relativity, but was ill-adapted
to the binary pulsar case because it presumed that the gravitational
fields involved were very weak. Nevertheless a somewhat similar, but
much less general (focusing as it did upon the case of gravitational
radiation emission)
parameterization was established which facilitated the theory
testing aspect of Weisberg and Taylor's 1981 paper. To quote from
Clifford Will's paper on the subject (1977)
\begin{quotation}
Because of the complexity of many alternative theories of gravitation
beyond the post-Newtonian approximation, we have not attempted to
devise a general formulation analogous to the PPN framework beyond
writing equation (2) with arbitrary parameters. However, we can provide
a general description of the method used to arrive at equation (2),
emphasizing those features that are common to the theories being studied.
\end{quotation}

So given the existence of a pidgin to create a trading zone between
astronomers (and others) interested in doing theory testing and 
gravitational theorists, why did the astronomers shrink from commenting
directly on the quadrupole formula itself? One obvious answer is that
the pidgin was not designed to facilitate such a conversation. It permitted
comparisons between calculations derived from different theories. It was
not designed for the more complex and open-ended task of critiquing
subtle details of such calculations. Another
answer is that the barriers were as much social as linguistic (the
two must obviously be linked). The astronomers felt they lacked
the social standing to weigh in on a question which obviously fell
within the purview of the theorists. Because the controversy over
which calculation within a given theory was the correct one depended
on subtle judgments, it naturally required the expertise of the
practicing theorists. This is precisely the meaning of the 
Theoreticians' Regress, that it depends on subtleties of expert
judgment and not on some closed algorithmic model of how to perform
a calculation.

\section{Skeptics' Dilemma}

I have argued that the closing of debate in the quadrupole 
formula controversy
occurred at least partly because of the quickening effect caused by the
binary pulsar increasing the importance of the controversy.
At the same time, the lifetime of the controversy, once the binary
pulsar data became available, was greatly constrained by the existence
of experimental data which bore directly on the topic at issue. For
the theoretical controversy to continue indefinitely, there would have
to have been a significant effort to contest either the experimental
evidence or the interpretation of it. The fact that there was no such
significant attack on the ruling interpretation of the binary pulsar 
data certainly limited the lifetime of the controversy, so it is 
interesting to look at the reaction of the skeptics to the work of
Taylor and his collaborators.

In any problem of orbital mechanics there are many mechanisms
which might account for all or part of an observed change in orbital
period. That even the most famous agreements between theory and
observation can be challenged in this way is shown by the saga of
Robert Dicke's efforts to measure the oblateness of the Sun (the degree
to which its shape departs from a perfect sphere). Dicke had
pointed out that if the Solar oblateness turned out to be significantly
different from zero, its gravitational influence on the orbit of
Mercury would throw out the close agreement between the prediction
of General Relativity and the observed perihelion advance of the
planet Mercury (Dicke and Goldenberg, 1967). As with the case of 
the Mercury Perihelion, the binary pulsar data seemed particularly
impressive because it agreed with the prediction of the quadrupole
formula with little or no need to take into account of other factors.
The interpretation was that the system was very ``clean.'' The
corollary to this, naturally, is that any evidence that the system
was not so clean would throw out the agreement.
Given this opening to challenge the {\it interpretation} of the
binary pulsar data, it is interesting that the gravitational
wave skeptics were not involved in proposing alternative mechanisms.

Certainly there were those who considered it, amongst them Peter
Havas and, very likely, his former student Arnold Rosenblum. They
were to the fore in demanding that the observations not be accounted
a successful test of general relativity given that (in their opinion) 
the quadrupole formula had
not been shown to a valid prediction of that theory. Joe Taylor 
recalls that certain people were particular about this question
of terminology (interview,
2nd May, 2008).

\begin{quotation}
Well, let me think, the people who kept bugging me about it, so to speak, 
were Peter Havas, Fred Cooperstock and Arnold Rosenblum. Arnold 
bugged me about it a lot. Anyway, they just kept saying `Look, even though 
you have an experimental number now, we're not even sure what the theoretical 
number is and you can't go around saying that you've confirmed something.' 
So I tried to remain outside of the argument, letting the 
theorists fight it out until they all ... persuaded one another. So that 
seemed to be the best thing for me to do and we were simply concerned with
getting an experimental result that we were happy with.
\end{quotation}

The alternative scenarios to the gravitational wave interpretation
were actually put forward in print, but generally
not by the skeptics. This may have been because the skeptics
found themselves in a similar position to the experimenters. They had
a vested interest in the debate, 
but lacked the special expertise which would have permitted
them to comment. Likely dissipative mechanisms (or even non-dissipative
ones) fell within the purview of astrophysics rather than relativity,
and were explored and commented upon by astrophysicists rather than
relativists. 

The most important issues which had to be dealt with in demonstrating
that the observed decay agreed with the quadrupole formula prediction
was the nature of the unseen companion in the system, and the relative
acceleration of the binary pulsar to our solar system. If the unseen
companion was a sufficiently compact object, like another neutron 
star (which is now firmly believed to be the case) then it would undergo
little deformation as a result of the visible pulsar's tidal effect.
But if it was a normal star, it would develop a marked oblateness
which would in turn create a perturbation in the orbit of the pulsar
(a tidal friction-like effect) which would be difficult, except over
longer timescales, to distinguish from the orbital decay due to
radiation damping. 
Effects of this type would, however, have affected other measurements
made in the system, and with time the experimenters became convinced
that the system was extraordinarily clean. As Taylor and McCulloch stated
in their paper from the Texas Symposium (1980)

\begin{quotation}
If one were given the task of designing an ideal machine for testing
gravitation theories, the result might be a system rather similar to
PSR1913+16; an accurate clock of large mass and small size, moving at
high speed in an eccentric orbit around a similar object located in
otherwise empty space. To be sure, one would place the system somewhat
closer to the Earth than $\sim$ 5 kpc, or which arrange for a more 
powerful transmitter to convey the clock pulses to terrestrial telescopes;
but we cannot expect Nature to be concerned with the inadequacies of
our instrumentation!
\end{quotation}

This sense of wonder at the sheer serendipity of coming across such
a system (many relativity theorists had sworn for decades that no 
system would ever be found in which gravitational wave effects would
be measurable) was brought into focus for me after the more recent
discovery of the ``double pulsar'' a system with an even closer orbit
than the original binary pulsar, in which both pulsars are visible
from Earth. I have heard this system referred to as ``a relativistic
astrophysicist's wet dream.''

Taylor and McCulloch's comment illustrates the three main technical
challenges in creating a match between theory and experiment for this
system. First, the system must be in empty space. The presence of 
instellar gas, for instance, would certainly alter the orbit of the
system with time, as a result of dynamical friction. A related issue
would be if the pulsars themselves were blowing off material at a
significant rate, in
which case the mass loss would affect the orbital motion. Secondly, as we 
have seen, both objects must be
compact objects, such as neutron stars, so that perturbations due to
the failure of the bodies to behave as point sources can be ignored. As
a corollary to this, if the system contained a third massive object,
this would obviously also affect the orbit of the two known components.
Finally, the object should be close to us, not only for reasons of
detection, but because a more distant object is in a more different orbit
around the center of the galaxy and would be accelerating more
strongly with respect to us here on Earth (for a list of references
and discussion of a number of these issues, see Damour and Taylor 1991).

It is a well known result of special relativity that systems which
are in inertial motion with respect to each other have clocks which
run at different rates. If the systems are accelerating with respect
to each other, then their respective clocks will alter, with time, in their
relative rates of running. Since the solar system and the binary pulsar
system are in different orbits around the galactic center they
are not in the same inertial frame with each other. Accordingly the
sensitive timing which is required to measure the orbital damping
effect is also capable of measuring the relative accelerations of
these two systems. In so far as doubt persisted about the validity
of the quadrupole formula, this was a bad thing. Indeed, at one point
during the 1980s, it did happen that the analysis of measurements
of the binary pulsar did fall out of agreement with the quadrupole
formula, by a much smaller amount than had been at issue in the 
earlier theoretical debate (in so far as that debate had ever been
completely quantified). A close analysis of the relativistic theory of timing
between the two systems, carried out by Taylor in collaboration
with Thibault Damour, showed that the discrepancy could be explained
on the basis of fully accounting for the timing issues (Damour and
Taylor, 1991). 

Ultimately, as Taylor recalls, the situation reached the point where,
if one {\it assumed} the validity of the quadrupole formula, one
could make an accurate determination of the position of the binary
pulsar in the galaxy, based on its relative acceleration. This measurement
was more
accurate than was possible by other methods at that time. This makes as good a
moment as any to mark the end of the quadrupole formula controversy.
When a prediction turns from a thing to be tested to a tool to be
used, the debate is surely closed (and this, of course, goes some way
to explain the impatience of non-skeptics to achieve that moment of
closure). It is a mark of the importance of
the controversy that the measurement of the distance to the galactic
center which could have been provided by the binary pulsar data
never became a canonical one, though it is in agreement with subsequent
measurements using other techniques.

As Damour and Taylor put it in 1991

\begin{quotation}
If we assume that the standard general relativistic framework ... is
valid we see that, in a few years, the measurement of 
$\dot{P}_{b}^\mathrm{obs}$ [the rate of
decay of the binary pulsar's orbit] can be turned into a measurement
of ... the galactic constants $R_o$ [the distance form the Solar System
to the Galactic center] and $v_o$ [the speed of galactic rotation at
about the center at the position of the solar system] (especially
$v_o$, which presently contributes the biggest uncertainty). Such
a ``pulsar timing'' measurement of $v_o$ would be free from many of
the astrophysical uncertainties that have plagued other determinations.
\end{quotation}

Since the Taylor-Hulse discovery, 
subsequent binary pulsars have been found where the relative
acceleration of the two systems does not permit a particularly
accurate determination of the rate of orbital damping. Had the controversy
persisted so far this might have provided some opening for
skeptics. However the discovery of the double pulsar in 2003,
a system in which both pulsars are oriented so that both their radio
beams are visible from the Earth, has provided a system with
even stronger orbital damping than the original binary pulsars,
whose results are in agreement with it.

How much interpretive flexibility was there for skeptics to
continue the controversy? Did the skeptics largely abandon the
fight because, as Franklin would have it, they were rational
actors or, as Collins would have it, they had run out of
sociological space in which to continue the argument? I suspect
both considerations played a role. A rational actor will certainly
take sociological considerations into account when determining whether
to continue a debate. Most physicists do not wish to face social
ostracism, even in a cause they believe to be right. At the
same time any social constructivist will agree that the ruling out
of certain arguments as work in the field progresses, the limitations
placed on interpretative flexibility in the ebb and flow of debate,
can tax the ingenuity of even the most stubborn
skeptics to the point at which they give up the struggle. The
social struggle can become
unequal in a double sense, in that they are both outnumbered
and outmaneuvered by their opponents. Whether the maneuvering
was all in vain, given the inevitable verdict of nature is, of course, 
an interesting question,
but not one that is trivial to answer by the historian's method.

That skeptics {\it considered} continuing the battle is clear enough.
Although Fred Cooperstock did retire from the fray for a
decade or so after the mid-eighties, he subsequently put forward
a new argument that gravitational waves would not propagate energy
through empty space. The failure, to date, of the new generation
of gravitational wave detectors like LIGO, to detect gravitational
waves passing by the Earth, has provided a new opening for skeptics
like Cooperstock. He and others now put forward arguments that
the existing theory is correct for {\it sources} like the binary
pulsar, but fails for {\it detectors} like LIGO, thus explaining
why we see evidence for gravitational waves in these source systems,
but cannot, as yet, detect them.\footnote{We cannot hope, with current
technology, to detect the gravitational waves emitted by the known
binary pulsar systems. It is only when such systems reach their terminal
point and spiral into each other and merge that Earth-based detectors
can hope to observe them.} The specifics of these new skeptical
arguments vary widely. Some come from professional physicists like
Cooperstock, others come from amateurs who focus on the sheer expense
of the detectors which, they claim, can never succeed in detecting
anything.\footnote{A sample of modern gravitational wave skepticism
is given by the following references: 
Cooperstock 1992, Bel 1996, Aldrovandi et al 2008, and, for the
non-professional viewpoint, see the webpage
http://www.god-does-not-play-dice.net/Szabados.html\#SBG, accessed
on April 24th, 2009} 
 
Peter Havas, when I interviewed him in 1995, certainly spoke of
the openings he believed had existed, at least for a time, for an 
attack on the standard
interpretation of the pulsar timing results. He still entertained
significant doubts about the consensus which had emerged at that
time. Joe Taylor reports that Havas, and
his student Arnold Rosenblum, did ask to see some of the data 
and that he sent them a magnetic tape containing some (private
communication). When he
asked them a year later whether they had made progress they
indicated that they had been distracted by other problems.
Nevertheless, a search for
Arnold Rosenblum's papers on the SAO/NASA Astrophysics Data System server
shows that, from the mid-eighties, after
several years spent on his calculations of gravitational wave
emission that did not agree with the quadrupole formula, he then
devoted a number of papers to the problems of relativistic timing
in orbital and binary systems. Although none of this series of papers
referred directly to the binary pulsar, they are strongly suggestive
that he had spent a considerable amount of time thinking about this
issue, leading him into that field.\footnote{Arnold Rosenblum
died tragically young in 1991 (Cohen, Havas and Lind 1991).}

Therefore we can say that the skeptics considered a foray against
the conventional interpretation of the binary pulsar data, but
decided against it. One can say that the physics of the
situation obliged them to react this way,
in that they felt they could not overturn the hard empirical
evidence provided by the binary pulsar data. But one can also say
there were sociological reasons. They were not in a position to
do their own experiment to challenge the data, because they lacked
the standing in that field which would have permitted them to enter
it with any hope of success. For starters they would never have been
granted time on a radio telescope to do their own measurements of this
system (one group of astronomers did do some independent timing
measurements of the binary pulsar, guided by data supplied by Taylor,
and concluded that Taylor and his collaborators were correct in their
results on the orbital decay, see Boriakoff et al 1982).
Even worse, in so far as the interpretation
of the data could be challenged by theorists, it was by
astrophysicists with experience in the study of stellar binaries
and pulsars, not by relativists experienced in gravitational waves.
Thus from a professional point of view the skeptics were in a 
double bind which, combined with their increasing isolation within
their own community, as the debate moved towards a final resolution,
prevented any kind of continuation of the public debate. Whatever
private doubts were held by  a few theorists about the reliability of
the existing calculations, the empirical result was regarded as
beyond dispute. The final option open to the skeptics, arguing
that Taylor had simply got it wrong, was undoubtedly not entertained
because of the outstanding reputation which Taylor enjoyed within
the astrophysics community for his careful and painstaking work.

This whole issue of replicating experimental work lies at the
heart of the philosophical controversy between Collins and Franklin
alluded to earlier. In his papers on the Weber controversy Collins
showed how problematic was the use of replication to close a debate.
The difficulty lies in the fact that the details of experiments
must necessarily differ, and these differences generally provide
ammunition for one side or the other. In addition, what is sauce
for the goose is sauce for the gander. If someone who replicates
an experiment charges that the original experimenter got it wrong,
the same charge can always be thrown back in their own teeth. 
Franklin responded that physicists could still rationally and
(hopefully) dispassionately decide between these competing claims.
From the sociologists' point of view the issue casts interesting
light on the nature of {\it expertise} and how it is recognized
and evaluated by fellow experts. From the sociologists standpoint,
the physicist, as a rational actor, must make a series of
social judgments over the course of a controversy
in evaluating his colleagues' expertise and the consequent reliability 
of their work.  

In the case of the binary pulsar replication demanded access to
radio telescope time to look at the same system or, better, the
discovery of an independent system. But, as we have seen, subsequent
systems were often not as ideal for this experiment as the original.
Not until the discovery of the double pulsar can we be said to have
a fully comparable replication of the original, so one can certainly
speculate that there may have been some scope for further controversy
in the decades between 1980 and the early years of the twentieth
century, had there been sufficient sociological space to support
such a debate. But while logical space for disputation may have 
remained, the skeptics had run out of sociological space.
Indeed, there is every reason to believe that the
field of gravitational wave physics could ill afford to permit such
a controversy to linger for that amount of time, lest it put its
own disciplinary space at risk.

\section{Theory Testing}

An interesting feature of the early papers on the orbital
decay measurements of the binary pulsar is the focus on theory
testing. In the 1981 paper by Weisberg and Taylor much of the
paper is devoted to discussion of the predictions of a variety
of alternative theories of gravity which were falsified by
the measurements. The best known of these theories was the 
bi-metric theory of Nathan Rosen, a longtime skeptic of gravitational
waves (Rosen, 1940). 
Rosen's theory had been shown to make a prediction of
anti-damping for binaries emitting gravitational waves 
(Will and Eardley, 1977). The
waves would carry away negative energy from the system, leaving
it more energetic than before, and thus permitting the orbiting
bodies to spiral away from each other. As Will and Eardley acknowledged
in their paper (p. L92)
\begin{quotation}
Some might thereby argue that the theory should be ruled out on
theoretical grounds alone.
\end{quotation}
But the theory was of particular interest to theory testers because
it agreed with the predictions of general relativity in the post-Newtonian
limit. Accordingly it was one of a handful of theories which had
survived all early solar-systems tests of gravitation theories. This
placed it in a special category of theories which could play a useful
role as a foil for theory testing with the binary pulsar, even if
calculations such as these were beginnings to show that the theory
had troubling pathologies. 

The emphasis on theory testing in early papers on the orbital
decay seems odd when these papers are read today. This work is famous,
but certainly not famous for invalidating the bi-metric
theory of Nathan Rosen. What seems particularly odd is that the
prediction of Rosen's theory (and other theories) which it invalidated
appeared paradoxical. As was openly acknowledged, no one would give
any credence to these predictions even without an experiment to
falsify them. The theory had few, if any, proponents by this stage.
It's certainly true that
we cannot say here that we have a direct confrontation between theories,
in any symmetric sense. While Rosen's theory could have been 
falsified by the results, it could not have been confirmed. Had
Taylor and colleagues encountered a result in agreement with Rosen's
theory's prediction, all sorts of other mechanisms would have
been proposed to explain it before Rosen's. Perhaps, given enough
supporting evidence (several other systems behaving the same way)
Rosen's theory would have been accepted, but it would have been a 
long hard road. 

Nevertheless, no matter how little credibility a
theory has, experimenters still find it satisfying to have a definite
prediction they can test. One must not be too inclined to overlook the
obvious motivation. Indeed, a crazy prediction of a reasonable theory,
as long as it is a definite prediction, may be a godsend to an experimentalist.
After all, falsifying such a prediction is likely to be seen as good,
worthwhile work by colleagues, and yet it will also be
uncontroversial and easily accepted by the community. Furthermore
a theory like Rosen's, with its odd prediction, plays a useful role
in the framing of experimental results as theory testing. It is a straw house
theory, in the sense that it is rather like
the pig who built his house out of straw. The main purpose of the research
is to show that general relativity has been validated. Therefore general relativity is like the house 
of bricks which does not fall down to the huffing and puffing of the big
bad wolf. But the story of the one little pig is rarely satisfying to an
audience. In order to appreciate the part about the pig who survived, we must
first learn about his brothers who were not so lucky. The foolish pigs
who built their houses of straw and sticks are perhaps all the more
welcome, from a narrative standpoint, if the brick house is the subject
of controversy. Doubts have been voiced as to whether the brick house
really was built by the third pig. Perhaps, say his detractors, said pig 
has been given too
much credit. How much easier it is to talk about the first two pigs.
At least no one is trying to claim credit for their edifices!\footnote{
parts of this section are based on an unpublished paper by the author
and Harry Collins.}

Lest anyone think that it is normal to find theories with strange predictions
waiting to be falsified, one must give Rosen some due credit here for not
contesting the calculation which set up this scenario. Clifford Will,
a leading figure amongst theorists interested in theory testing experiments,
who was the chief architect of the parameterized post-Newtonian
scheme mentioned previously, was very active in producing the calculations
which provided predictions from alternative theories. Note the very fact
that the authors of the theories were not doing these calculations themselves
suggests that we are not dealing with theories which have proper
communities of advocates behind them.
Will notes that it is relatively unusual to find that the author of a
new theory will agree with a calculation which shows that the theory
makes a prediction that is highly likely to be falsified by
experiment. Generally the process, typically of many scientific
controversies, can be almost open-ended (Interview conducted by
the author at Washington University, St. Louis, 2nd March, 1999).

\begin{quotation}
It can be, and it rarely reaches a conclusion. The only time I know .. 
and I don't get involved in this all that much. I mean, I don't 
grab theories out of the literature and analyze them. It's kind of a 
hopeless and not a terribly rewarding task. But my experience has been that 
it takes a long time because the people who propose it always try 
to wriggle out of it. But there are only two cases that I know of where 
it has actually come to a conclusion whereby the person said,‘`yes I agree, 
this theory is wrong' and one was Rosen himself. Because when we did this 
work on the binary pulsar and showed that Rosen's theory disagreed with 
the observations, in fact I was giving a talk in Haifa shortly after that 
and gave this lecture and said that Rosen's theory is wrong and at the end 
of the lecture Rosen stood up and said `yep, I agree with you, it's wrong, 
... but I have a new theory'’ a rather different theory which he then went 
on to argue had nice properties and agreed with all the experiments.
\end{quotation}

It is worth noting here that Will's work showed that Rosen's theory
predicted negative energy wave emission only in the case of dipole
gravitational waves. This in itself was a departure from standard
general relativity theory, since dipole gravitational waves do not
exist in this theory. Even in Rosen's theory dipole radiation would
not be emitted for a binary system consisting of two identical
pulsars. Since it was gradually shown that the binary
pulsar and its companion are fairly similar, Weisberg and Taylor (1981)
found it necessary to calculate the quadrupole prediction of Rosen's
theory (and certain other theories which were also falsified
by their work) themselves, drawing upon Will's framework. The calculation
showed that Rosen's theory predicted negative energy waves even
in the quadrupole case. 

But if Rosen's theory predicted an unphysical result, wouldn't it
have been discarded even if the binary pulsar hadn't been found
to falsify it?
To quote Cliff Will again (interview, 2nd March, 1999)

\begin{quotation}
In a case like that it really depends on your point of view. Some people 
would have argued that just having anti-damping, negative energy 
flux would make that a bad theory right off the bat and you would 
throw it away without further ado. So my attitude is slightly 
more phenomenological then that. I'm willing to say that it looks strange 
to me but let's compare it with observations and, of course, there the 
comparison is easy because we see damping and not anti-damping
and so it really is wiped out. But some people would just say on theoretical 
grounds, `that theory's dead.'” 
\end{quotation}

The context here is particularly important. Physicists interested in
gravitational waves were used to having no experiments at all. Once an
experiment had, at last, become available, they wanted to put it to
every kind of use they could, and theory testing was the most established
role for experiment in the general relativity community. Most of the
work in this field which had some prestige in the wider physics community
was of the theory testing variety, such as the British 1919 eclipse
expedition, the Pound-Rebka experiment, the perihelion advance of Mercury
and the Shapiro time-delay measurement. The limitation of all of these
experiments, as far as theory-testing goes, was that they were all
``solar-system'' tests limited to weak gravitational fields. As such,
some theories, and Rosen's was a leading example, could not be distinguished
from general relativity by these tests. It was certainly natural for
those involved with the binary pulsar to anticipate that its significance
would lie largely in the fact that it was the first strong-field test
of general relativity and its rivals. In fact, such was the significance
of the discovery of evidence of the existence of gravitational waves,
that this quickly came to dominate everything else. As such, the
falsification of Rosen's theory seems almost quaint today, compared
to the importance of the verification of the general relativistic
quadrupole formula.

\section{Conclusions}

It is now time for me to examine my own place within a controversial
field, in analogy with my study of the astrophysicists struggling
to interpret the binary pulsar data. I find myself trying to
interpret their struggles in the context of the competing theories
of social constructivism and rival philosophies which insist
on the normative standing of experiment, permitting it a special
status in deciding scientific controversies. It is in this sense
that the binary pulsar story may be said to be a potentially controversial
case study from the science studies standpoint.

Is there a sense in which my work can decide between
these competing theories? Unfortunately the answer appears to be
no. Perhaps this is fortunate, since I mentioned at the outside
that I am not sure I want to place myself squarely in the cross hairs
of this particular controversy. The problem is that the predictions
of the two theories do not significantly differ from each other in
this case.
At the resolution provided by my study, there does not seem to exist
a possibility of deciding between them. Collins would say that
Kennefick has extended his notion of the Experimenters' Regress
onto the Theoretical side in a way that seems natural and useful.
He would say it is not at all surprising that the theorists' seeking
a way out of the regress, should turn to an outside expertise, in the
form of experimenters, to find a resolution. It is just the inverse
of the way in which experimenters, seeking a way out of their
regress, might appeal to theory in order to decide between 
competing experimental results. But of course Franklin is perfectly
happy for theorists to let their debates be decided by experiment.
It fits in completely with his normative picture of experiment
as the decisive factor in such disputes. Thus each side is likely
to be happy with the basic story I've outlined. Even more problematic
is the way in which the philosophical debate does not necessarily
permit a clear distinction to be drawn in the behavior of the
scientists involved. Even if we concede that the protagonists were
more willing to settle the issue based on what the experimenters
said, and were relatively unwilling to challenge what the experimenters
said, this could be explained by the sociologists as simply a feature
of the society under examination. Theorists, by the rules of the
game, have a relatively limited (but definitely non-zero)
liberty to challenge the expertise
of experimenters. When they do so, they must do so from a position
of strength, and the entire history of the controversy shows that
the skeptics were already in a position of weakness by the time
the binary pulsar data came along. Indeed we have noted that
a paradoxical effect of the experimenters' arrival on the scene
was to breathe new life into the controversy, effectively giving
new oxygen to the skeptics, even as it forced them to consume
more oxygen in the exertion of defending their position.

Recall that the substance of the debate between
the physicist/philosopher Franklin
and the sociologist Collins was the problem of replication,
and how one can tell whether an experiment has been performed
correctly. To some extent it boils
down to the question of how scientists deal
with the possibility that Joe Taylor and colleagues might simply
have gotten it wrong. This is especially noteworthy in this case,
because for a considerable time there was no confirming experiment.
The answer is that most people were impressed with Taylor and felt
they had every reason to trust his work. This is a profoundly
sociological issue obviously. Even if one believes that Taylor is
correct, one has not actually done all the work he has done to
convince himself. One assumes that he has done a proficient job,
especially if one has had reason to believe that other work he
has done has been very reliable. In short, at least one important aspect
of judging the work of fellow physicists derives from our ability
to judge their standing in the community and to assess their expertise
from social encounters. Note that only if we are physicists ourselves
are we likely to have much success in making this kind of judgment.
Collins' and Franklin's debate concerns
(in part) the question of how much importance one should place on this
aspect of the reception of scientific work. 

For what it is worth, my own view is that at bottom physicists
are simply doing what humans usually do and applying a basic version
of the principle of induction. If an experiment is replicated $n$
times and always produces the same result, then the $n+1$th replication
will produce the same result. It makes sense for physicists to
infer that this is because reality is determining the outcome of the
experiment. For a sociologist, it makes more sense to assume that
if $n$ experimenters sharing a similar expertise perform an experiment
the same way, then the $n+1$th expert will perform it the same way
and also produce the same result. The inference is different, depending
on the academic interests of the scholars involved, but the basic
principle is the same. It seems to me that the argument between some 
sociologists
and some philosophers on this topic is similar to the old dispute between
realists and empiricists. Philosophers are saying that science is only
possible because scientists are engaged in studying a real entity, the
laws of nature. In this case, the sociologists are in the role
of the empiricists, insisting that sense impressions are the only reality
and observing that much of what passes for sense perception in modern
science is what scientists hear from other scientists as scientific
knowledge passes through a series of social networks.
I doubt that I can decide a debate between realism
and empiricism.\footnote{Another question I cannot answer is if
$n$ historians study the same historical
episode, can we rely on the $n+1$th historian reaching the same
conclusions? Can I do so if $n=1$? 
Is there any sense in which historical micro-studies of this
kind can be compared to real science? 
Is there a Historians' Regress related to the problem of When History
Ends, just as the Experimenters' Regress relates to the problem
of When Experiments End? 
The phrase ``When History Ends'' may seem millennial in tone
but note the aptness of the word Apocalypse
which means, ``the lifting of the veil,'' which
is entirely what the historian is trying to so. Just as the
radio astronomer does in continuing his timing measurements over
longer periods to greater degrees or precision, or as the theorist
does in delving to higher orders in an approximation scheme,
so the historian burrows down more deeply in
a micro-study. 
But in historical analysis we should be careful
to practice Interpretational Frugality, a sort of inverse 
form of Occam's Razor. We may multiply entities if it is in the service
of keeping our feet grounded in the local. Not all morals are generally
applicable.  In the words of Bart Simpson, sometimes
there is no moral, ``just a bunch of stuff that happened.''}

The moral of this story, it now seems to me, is that science works. Does
this mean that the story I am telling is an argument in favor of realism? 
It's certainly not an argument against realism and its true that the strong
realist would say that science has to work because the objective nature of
reality constantly obtrudes on experimental work of all kinds. But if we
adopt the position of the strong program of the sociologists, we must
work rather harder to explain how it is that scientists ``manufacture
consent.'' This question is of interest even to the realist, since history
certainly tells us that people are sometimes wrong about the laws of nature.
In this imperfect world, if scientists are able to reach agreement amongst
themselves, we can announce that science works. This sounds like a very
global moral, but its true significance is local. The relativity community,
I believe, had quite a lot at stake in this debate. They had to show that
{\it their} science worked, that they as a community could do science
which worked. The binary pulsar therefore
played a key role in showing that relativity as a field, and relativists
as a community, could work as a functioning branch of science, that
relativists were competent, and not dopes. From
the point of view of the relativists, it would have mattered little
whether their behavior was viewed as that of incompetent, irrational physicists
who refused to accept the obvious fact that gravitational waves existed,
or as that of needlessly fractious and insufficiently socialized actors 
unable to crystallize a core group amongst themselves in order to facilitate
normal scientific behavior. What they had to do was demonstrate that
they were a healthy branch of physics, a postulate
which Feynman, sitting in the Grand Hotel, Warsaw in 1962, would have
doubted. The imagery confronting Feynman as he set in a hotel
restaurant and contemplated this dysfunctional field, writing the script
for a possible Fellini film, ``126 Dopes'' has been replaced by
the gleaming, high tech, ultra-precise big science of LIGO, and by the 
confidence of funders in pouring money
into the construction of large gravitational wave detectors. Since no
one can know for certain whether gravitational waves will ultimately be
detected by these devices, the process by which the more confident 
decided to begin ignoring the anxieties of the more cautious is an
interesting one regardless of whether we believe those cautious skeptics
were irrational dopes or sensible social actors.

\section{Acknowledgments}
I would like to thank Joseph Taylor, Clifford Will, Thibault Damour, 
Joel Weisberg and the late Peter Havas all of whom permitted me to interview 
them for the research which gave rise to this paper. All of the interviews,
except the one with Peter Havas, were recorded. Both Harry Collins and
Allan Franklin discussed some of the issues bearing on this paper with
me many times, and aspects of it are based on an unpublished 
draft of a paper written by Collins and I. I would like to thank both
of them for their help and inspiration on this work. Diana Buchwald and
Kip Thorne both helped me far more than I can recall in the early
stages of this work, and I would also like to thank David Rowe for
his giving me the chance to finally turn it into a paper, and for his
patience waiting for it to be finished.

\section{Bibliography}
\bigskip
\medskip
Aldrovandi, R, Pereira, J. G., da Rocha, Roldao and Vu, H. K. (2008).
``Nonlinear Gravitational Waves: Their Form and Effects.'' arXiv:0809.2911v1

\medskip
Baade, Walter and Zwicky, Fritz (1933). ``Remarks on Super-Novae and Cosmic
Rays'' {\it Physical Review} 46: 76-77.

\medskip
Bel, Luis (1996). ``Static Elastic Deformations in General Relativity''
electronic preprint gr-qc/9609045 from the archive http://xxx.lanl.gov

\medskip
Boriakoff, Valentin, Ferguson, Dale C., Haugan, Mark P., Terzian, Yervant
and Teukolsky, Saul A.
(1982). ``Timing Observations of the Binary Pulsar PSR 1913+16.'' {\it
The Astrophysical Journal} 261: L97-L101.

\medskip
Cohen, Jeffrey M., Havas, Peter and Lind, V. Gordon (1991). ``Arnold Rosenblum.'' {\it Physics Today} 45: 81. Another obituary of Rosenblum appeared in 
the {\it New York Times} of January 7, 1991.

\medskip
Collins, Harry M. (1994). ``A Strong Confirmation of the Experimenters'
Regress,'' {\it Studies in History and Philosophy of Science Part A}
25: 493-503.

\medskip
Collins, Harry M. (2004). {\it Gravity's Shadow} Chicago: University of Chicago
Press.

\medskip
Collins, Harry M. (2009). ``We cannot live by scepticism alone.'' {\it Nature}
458: 30-31.

\medskip
Collins, Harry M., Evans, Robert and Gorman, Mike (2007). ``Trading Zones and Interactional Expertise.'' {\it Studies in the History and Philosophy of Science A} 38: 657-666. 

\medskip
Cooperstock, Fred I. (1992) ``Energy Localization in General Relativity: A
New Hypothesis'', {\it Foundations of Physics}, {\bf 22}, 1011-1024.

\medskip
Damour, Thibault and Ruffini, R. (1974). ``Sur certaines v\'{e}rifications
nouvelles de la Relativit\'{e} g\'{e}n\'{e}rale rendues possibles par la
d\'{e}couverte d'un pulsar membre d'un syst\'{e}me binaire'' {\it Comptes
Rendu de l'Academie des Sciences de Paris, series A} 279: 971-973.

\medskip
Damour, Thibault and Taylor, Joseph H. (1991). ``On the Orbital Period Change
of the Binary Pulsar PSR 1913+16'' {\it The Astrophysical Journal} 366: 501-511.

\medskip
De Witt, C\'{e}cile M. (1957).
{\it Conference on the Role of Gravitation in Physics},
proceedings of conference at
Chapel Hill, North Carolina, January 18-23, 1957.
(Wright Air Development Center (WADC) technical
report 57-216, United States Air Force,
Wright-Patterson Air Force Base, Ohio). A supplement with an expanded
synopsis of Feynman's remarks was also distributed to participants (a copy
can be found, for example, in the Feynman papers at Caltech).

\medskip
Dicke, Robert H and Goldenberg, H. Mark (1967). ``Solar Oblateness and
General Relativity'' {\it Physical Review Letters} 18: 313-316.

\medskip
Dyson, Freeman (1963). ``Gravitational Machines'' in {\it Interstellar
Communications} ed. A.G.W. Cameron New York: W.A. Benjamin Inc. pp. 115-
120.

\medskip
Einstein, Albert (1916). ``N\"{a}herungsweise Integration der Feldgleichungen
der Gravitation'' {\it K\"{o}n-\newline
iglich Preussische Akademie der Wissenschaften Berlin,
Sitzungsberichte:} 688-696

\medskip
Einstein, Albert (1918).  ``\"{U}ber Gravitationswellen''
{\it K\"{o}niglich Preussische Akademie der Wissenschaften Berlin,
Sitzungsberichte:} 154-167

\medskip
Einstein, Albert and Rosen, Nathan (1937). ``On Gravitational Waves''
{\it Journal of the Franklin Institute}, {\bf 223}, 43-54.

\medskip
Feynman, Richard P. and Leighton, Ralph (1988).
{\it What do \underline{you} care what other people think?
Further adventures of a curious character}
(Norton, New York). Remark quoted appears
on pg. 91 of the Bantam paperback edition (New York, 1989).

\medskip
Franklin, Allan (1994). ``How to Avoid the Experimenters' Regress'' {\it
Studies in History and Philosophy of Science Part A} 25: 463-491.

\medskip
Galison, Peter (1997). {\it Image and Logic: A Material Culture of
Microphysics} Chicago: University of Chicago Press.

\medskip
Haensel, Pawel, Potekhin, Alexander Y. and Yakovlev, D. G. (2007). {\it Neutron
Stars 1: Equation of State and Structure} New York: Springer. 

\medskip
Havas, Peter (1973). ``Equations of Motion, Radiation Reaction,
and Gravitational Radiation'' in {\it Ondes et Radiation Gravitationelles}
proceedings of meeting, Paris, June, 1973 Paris: Editions du Centre National
de la recherche scientifique, pp. 383-392.

\medskip
Hulse, Russell (1997). ``The Discovery of the Binary Pulsar'' in 
{\it Nobel Lectures in Physics
1991-1995} ed. G\"{o}sta Ekspong Singapore: World Scientific.

\medskip
Hulse, R.A. and Taylor, J.H. (1975).
``Discovery of a Pulsar in a Binary System''
{\it Astrophysical Journal}, {\bf 195}, L51-L53.

\medskip
Kaiser, David (2009). ``Birth Cry of Image and Logic'' {\it Centaurus}
50: 166-167.

\medskip
Kennefick, Daniel (2007). {\it Traveling at the Speed of Thought: Einstein
and the Quest for Gravitational Waves} Princeton, New Jersey: Princeton 
University Press.

\medskip
Rosen, Nathan (1940). ``General Relativity and Flat Space I'' {\it Physical
Review} 57: 147-150.

\medskip
Royal Swedish Academy of Sciences (1993). Press Release announcing the
Nobel prize winners in Physics for 1993, issued 13 October, 1993 and
retrieved on the web at http://nobelprize.org/nobel\_prizes/physics/laureates/1993/press.html on Apr 21, 1993.

\medskip
Taylor, Joseph H. and McCulloch, P. M. (1980). ``Evidence for the Existence
of Gravitational Radiation from Measurements of the Binary Pulsar 1913+16.''
in {\it Proceedings of the Ninth Texas Symposium on Relativistic Astrophysics}
ed. J\"{u}rgen Ehlers, Judith Perry and Martin Walker, pp. 442-446 New York:
New York Academy of Sciences.

\medskip
Wagoner, Robert V. (1975). ``Test for the Existence of Gravitational Radiation''
{\it Astrophysical Journal} 196: L63-L65.

\medskip
Weisberg, Joel M. and Taylor, Joseph H.(1981). ``Gravitational Radiation from an
Orbiting Pulsar.'' {\it General Relativity and Gravitation} 13: 1-6.

\medskip
Will, Clifford M. (1977). ``Gravitational Radiation from Binary Systems
in Alternative Metric Theories of Gravity: Dipole Radiation and the Binary
Pulsar.'' {\it The Astrophysics Journal} 214: 826-839.

\medskip
Will, Clifford M. and Eardley, Doug M. (1977). ``Dipole Gravitational
Radiation in Rosen's theory of gravity - Observable effects in the
binary system PSR 1913+16.'' {\it The Astrophysical Journal} 212: L91-L94.

\end{document}